\documentclass[aps,prl,twocolumn,floats,showpacs,superscriptaddress]{revtex4-2}
\usepackage{graphicx,amsmath,amsfonts,amssymb,upgreek,txfonts,color}
\usepackage[colorlinks,linkcolor=blue,citecolor=blue,urlcolor=blue,breaklinks=true]{hyperref}
\usepackage{color,colortbl}
\definecolor{lightgreen}{rgb}{0.88,1,1}

\begin{document}
	
	\title{Quantum Magnetometry with Orientation beyond Steady-State Limits in Cavity-Magnon Systems}
	
	\author{Zheng Liu}
	\affiliation{School of Physics, Dalian University of Technology, Dalian 116024, People's Republic of China}
	
	\author{Ding-hui Xu}
	\affiliation{School of Physics, Dalian University of Technology, Dalian 116024, People's Republic of China}
	
	\author{Yi-jia Yang}
	\affiliation{School of Physics, Dalian University of Technology, Dalian 116024, People's Republic of China}
	
	\author{Yu-qiang Liu}
	\affiliation{School of Physics, Henan Normal University, Xinxiang 453007, China}
	
	\author{Chang-shui Yu}
	\email{ycs@dlut.edu.cn}
	\affiliation{School of Physics, Dalian University of Technology, Dalian 116024, People's Republic of China}
	
	\begin{abstract}
		We propose a transient vector quantum magnetometry protocol based on cavity-magnon systems. By exploiting finite-time dynamics initialized from a reservoir-engineered squeezed steady state, our scheme retains residual squeezing-induced quadrature noise reduction, which suppresses transient added noise and enhances the short-time signal-to-noise ratio beyond conventional unsqueezed steady-state limits. IQ demodulation of orthogonal cavity-output quadratures enables crosstalk-free reconstruction of all three components of a transient magnetic field, providing access to both its magnitude and orientation. This vector capability is relevant for short-lived magnetic phenomena such as pulsed spin excitations, magnetic textures, nanoscale current transients, and biomagnetic signals. In the long-time limit, we derive a closed-form stationary noise spectrum and identify the on-resonance noise-cancellation condition $g_{am}=\sqrt{\kappa_a\kappa_m}/2$ at which the cavity-added noise vanishes without strong coherent coupling. Injected squeezing further suppresses the cavity-added noise away from resonance, while an array of $N$ yttrium iron garnet spheres reduces the magnon-probe noise contribution by a factor of $1/N$. Our results establish cavity-magnon systems as a scalable platform for transient, vector-resolved quantum magnetometry.
	\end{abstract}
	
	\maketitle
	
	\emph{Introduction.}---Quantum sensing \cite{RevModPhys.89.035002} exploits squeezing, entanglement, and criticality to enable ultrasensitive detection of weak signals across a broad range of platforms \cite{PhysRevLett.45.75,PhysRevD.23.1693,PhysRevLett.132.020801,doi:10.1126/science.aaz9236,PhysRevLett.121.020402,PhysRevLett.126.010502,PhysRevLett.124.120504,10.1063/5.0208107,Chao:21}. Quantum-enhanced magnetometers based on optically pumped atoms \cite{PhysRevLett.127.193601}, NV centers \cite{RevModPhys.92.015004}, and superconducting quantum interference devices \cite{PhysRevLett.111.067202} have achieved remarkable sensitivities. Beyond detecting the magnitude of weak magnetic fields, many emerging applications require the reconstruction of transient vector signals whose orientation and strength both carry essential information, including pulsed spin excitations, nanoscale current transients, magnetic textures, hybrid quantum devices, and biomagnetic or materials-related signals \cite{PhysRevLett.127.233202,Meng2023,Wang2025,Herb2025,Casola2018,Liu2025,PhysRevLett.129.087701}. Hybrid cavity architectures---particularly optomechanical \cite{PhysRevLett.108.120801,PhysRevLett.125.147201,zhang,PhysRevLett.133.153601,LiOuLeiLiu+2021+2799+2832,Li:18} and cavity magnon-polariton systems \cite{BARBIERI2017135,PhysRevB.99.214415,PhysRevApplied.16.034036}---have emerged as promising candidates due to their compatibility with miniaturization, on-chip integration, and enhanced detection capabilities.
	
	Among these platforms, cavity-magnon systems employing yttrium iron garnet (YIG) spheres are particularly attractive for quantum information processing and sensing \cite{PhysRevLett.104.077202,PhysRevLett.120.057202,Zhang2015,PhysRevLett.121.203601,PhysRevA.99.021801,PhysRevLett.123.127202,PhysRevLett.111.127003,PhysRevLett.113.083603,PhysRevLett.113.156401,doi:10.1126/sciadv.1501286,PhysRevLett.121.137203,PhysRevLett.125.117701,ZARERAMESHTI20221,PhysRevA.105.033507,PhysRevA.108.043703,Zhang2024}. Their high spin density, long coherence time, and wide tunability render them well suited for quantum-enhanced magnetic sensing \cite{10.1063/5.0024369,PhysRevA.103.062605}. The Kittel mode \cite{PhysRev.110.836} couples strongly to microwave cavity photons via magnetic dipole interaction and interfaces with optical photons through Brillouin scattering in whispering-gallery modes \cite{PhysRevLett.117.123605}, enabling applications ranging from precision magnetometry to microwave-optical conversion. These features make cavity-magnon platforms especially appealing for detecting weak microwave-frequency magnetic signals in integrated quantum devices.
	
	Nevertheless, the sensitivity is ultimately limited by quantum noise imposed by the Heisenberg uncertainty principle \cite{RevModPhys.82.1155,PhysRevLett.105.123601,PhysRevX.2.031016}. Although recent noise-suppression strategies---such as time-modulated cavity-magnon coupling \cite{PhysRevA.103.062605}, dual-frequency modulation \cite{PhysRevA.109.023709}, anisotropy-induced nonlinearities \cite{mqdg-wvkk}, and two-photon pumping \cite{https://doi.org/10.1002/qute.202500608}---have significantly improved weak-field detection, most existing cavity-magnon magnetometers still rely on steady-state spectral readout. Consequently, several fundamental limitations persist: (i) steady-state measurements assume long interrogation times, thereby erasing useful initial-state information and excluding finite-time transient dynamics that are crucial for short-lived signals; (ii) most protocols probe only a single field projection or require stringent alignment between the sensing axis and the unknown magnetic field, limiting their use when the field direction is unknown or time dependent; and (iii) the absence of vectorial resolution prevents unambiguous determination of the magnetic-field direction, restricting the information that can be extracted from realistic magnetic signals.
	
	To overcome these limitations, we introduce a cavity-magnon sensing scheme that combines transient analysis with full three-axis detection, enabling crosstalk-free reconstruction of magnetic fields with arbitrary orientations and extending magnetometry beyond steady-state and single-axis operation. Our approach relies on quadrature-resolved detection of the cavity output, which naturally separates the transverse and longitudinal field components and allows independent reconstruction of all three Cartesian components. This capability is particularly relevant for short-lived and direction-dependent magnetic phenomena, where the field orientation is as important as its amplitude. We analyze the signal-to-noise ratio (SNR) in both transient and stationary regimes. In the transient regime, finite-time dynamics and an engineered initial state are explicitly incorporated, yielding a closed-form finite-time noise spectrum. We show that the residual squeezing-induced reduction of quadrature noise inherited from the engineered initial steady state can significantly enhance the short-time SNR beyond unsqueezed steady-state protocols, providing a distinct advantage for magnetic-field reconstruction and orientation estimation within finite interrogation times. In the stationary limit, we derive a closed-form steady-state spectrum and identify the on-resonance noise-cancellation condition $g_{am}=\sqrt{\kappa_a\kappa_m}/2$, under which the cavity-added noise vanishes without requiring strong coherent coupling, leaving the sensitivity limited by the magnon probe noise. Away from resonance, injected squeezing further suppresses the cavity-added noise and broadens the detection bandwidth. Finally, extending the scheme to $N$ YIG spheres produces a collective bright mode that reduces the magnon-probe noise contribution by a factor of $1/N$, offering a scalable route to enhanced sensitivity. These results establish cavity-magnon systems as a promising platform for transient, vector-resolved quantum magnetometry with potential relevance to spintronics, condensed-matter sensing, hybrid quantum technologies, and biomagnetic detection.
	
	\begin{figure}[htbp]
		\includegraphics[width=\columnwidth]{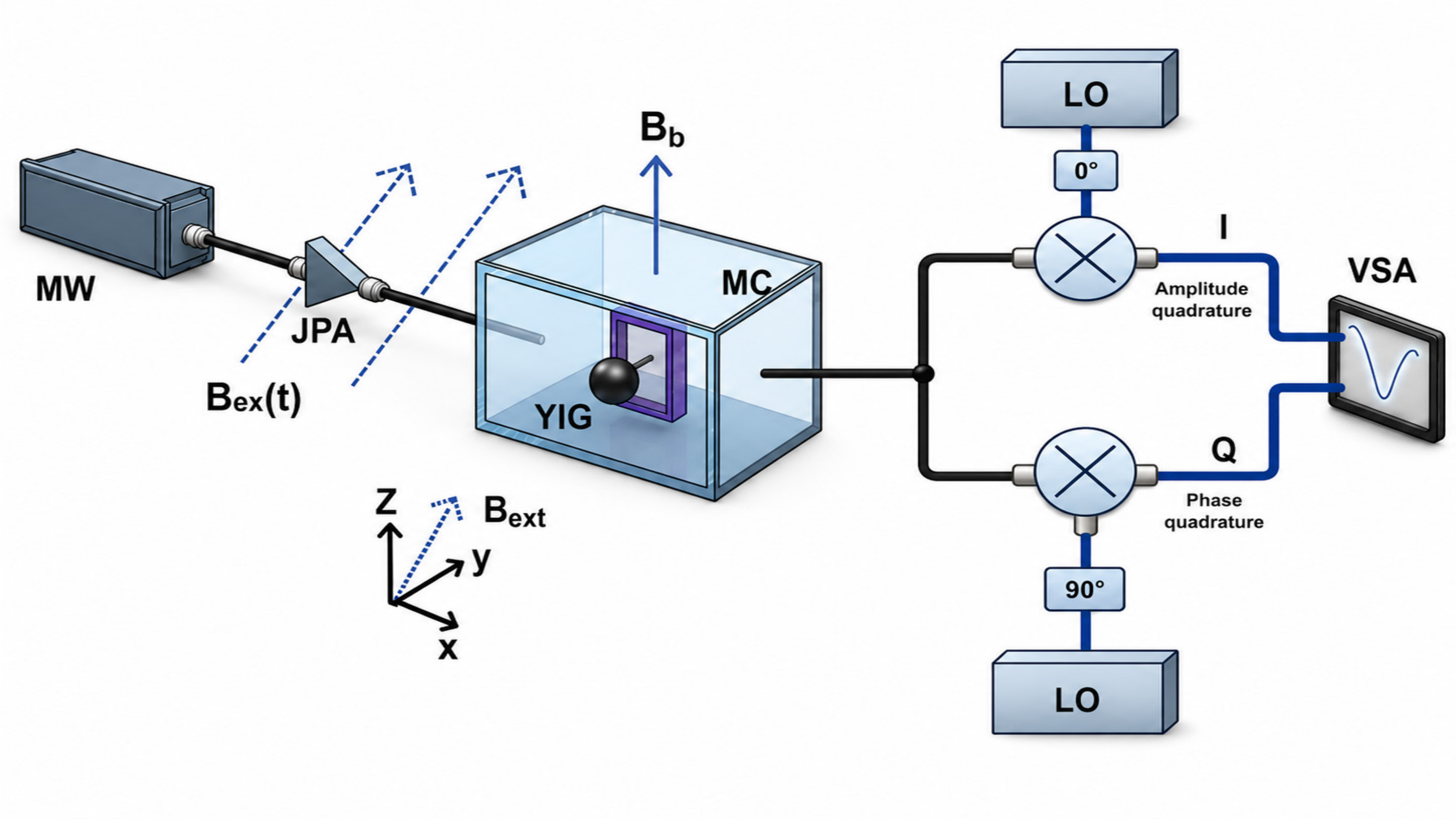}
		\caption{Schematic of the weak magnetic-field sensor. A JPA-generated squeezed microwave field probes a YIG sphere in a microwave cavity, where the bias field $B_b\hat{\mathbf{z}}$ tunes the magnon resonance and the transient field $\mathbf{B}_{\mathrm{ext}}(t)$ perturbs the magnon dynamics. The cavity output is IQ-demodulated to reconstruct the vector magnetic field.}
		\label{Fig1}
	\end{figure}
	
	\emph{Sensing scheme and dynamics.}---As illustrated in Fig.~\ref{Fig1}, a YIG sphere in a microwave cavity supports a collective magnon mode tuned by a static bias field. A weak magnetic field with arbitrary orientation perturbs the magnon dynamics and is mapped onto the cavity field through magnon-photon coupling. Squeezed microwave input and IQ demodulation of the output quadratures enable vector magnetic-field reconstruction. The system Hamiltonian is
	\begin{equation}
		\hat{H}
		=
		\hbar\omega_a\hat{n}_a
		+\hbar\omega_m\hat{n}_m
		+\hbar g_{am}
		\left(
		\hat{a}^\dagger\hat{m}
		+\hat{m}^\dagger\hat{a}
		\right)
		+\hat{H}_B ,
		\label{eq:H}
	\end{equation}
	where $\hat{a}$ and $\hat{m}$ are the cavity and magnon annihilation operators, $\hat n_a=\hat a^\dagger\hat a$ and $\hat n_m=\hat m^\dagger\hat m$, $\omega_m=\gamma B_b$, and we take $\omega_a=\omega_m$. The coupling strength is $g_{am}=\gamma B_a\sqrt{2sN_s}/2$, with $B_a=\sqrt{\hbar\omega_a\mu_0/(2V_a)}$, $s=5/2$, and $N_s$ the total spin number in the YIG sphere \cite{PhysRevLett.113.156401,doi:10.1126/sciadv.1501286}. The magnetic-field interaction is
	\begin{equation}
		\hat H_B
		=
		-\hbar\epsilon_B
		\left[
		B_-(t)\hat m+B_+(t)\hat m^\dagger
		\right]
		+\hbar\gamma B_z(t)\hat m^\dagger\hat m ,
		\label{eq:HB}
	\end{equation}
	where $\mathbf{B}(t)=[B_x(t),B_y(t),B_z(t)]$, $B_\pm(t)=B_x(t)\pm iB_y(t)$, and $\epsilon_B=\gamma\sqrt{sN_s/2}$ characterizes the magnon-field coupling. This term follows from the Zeeman interaction under the Holstein-Primakoff approximation in the small-excitation regime \cite{PhysRev.58.1098}; see Supplemental Material (SM), Sec.~I \cite{SM}. Thus, $B_x(t)$ and $B_y(t)$ act as coherent magnon drives, while $B_z(t)$ dispersively shifts the magnon frequency.
	
	We define the quadratures $\hat O_x=(\hat o^\dagger+\hat o)/\sqrt{2}$ and $\hat O_p=(\hat o-\hat o^\dagger)/(\sqrt{2}i)$, where $\hat o=\hat a,\hat m$. Including dissipation and quantum input fluctuations, the Heisenberg-Langevin equations in the interaction picture read \cite{PhysRevA.31.3761,PhysRevLett.46.1}
	\begin{align}
		\dot{\hat a}_x
		&=
		-\frac{\kappa_a}{2}\hat a_x
		+g_{am}\hat m_p
		+\sqrt{\kappa_a}\,\hat a_{\mathrm{in},x},
		\nonumber\\
		\dot{\hat a}_p
		&=
		-\frac{\kappa_a}{2}\hat a_p
		-g_{am}\hat m_x
		+\sqrt{\kappa_a}\,\hat a_{\mathrm{in},p},
		\nonumber\\
		\dot{\hat m}_x
		&=
		-\frac{\kappa_m}{2}\hat m_x
		+\Omega_z(t)\hat m_p
		+g_{am}\hat a_p
		-\sqrt{2}\epsilon_B
		\left[
		B_x(t)\sin(\omega_m t)
		+B_y(t)\cos(\omega_m t)
		\right]
		+\sqrt{\kappa_m}\,\hat m_{\mathrm{in},x},
		\nonumber\\
		\dot{\hat m}_p
		&=
		-\frac{\kappa_m}{2}\hat m_p
		-\Omega_z(t)\hat m_x
		-g_{am}\hat a_x
		+\sqrt{2}\epsilon_B
		\left[
		B_x(t)\cos(\omega_m t)
		-B_y(t)\sin(\omega_m t)
		\right]
		+\sqrt{\kappa_m}\,\hat m_{\mathrm{in},p}.
		\label{eq:HL}
	\end{align}
	Here, $\kappa_a$ and $\kappa_m$ are the cavity and magnon decay rates, $\hat a_{\mathrm{in},x(p)}$ and $\hat m_{\mathrm{in},x(p)}$ are the corresponding input noises, and $\Omega_z(t)=\gamma B_z(t)$. A squeezed thermal reservoir is injected to reduce one cavity quadrature noise \cite{PhysRevD.23.1693,Aasi2013}. The relevant input correlations are $\langle \hat a^{\rm in}(t)\hat a^{\rm in\dagger}(t')\rangle=(N_q+1)\delta(t-t')$ and $\langle \hat a^{\rm in}(t)\hat a^{\rm in}(t')\rangle=M_q\delta(t-t')$, with $N_q=\sinh^2r+\bar n_a\cosh 2r$ and $M_q=(2\bar n_a+1)e^{i\theta}\sinh r\cosh r$. Here, $r$ and $\theta$ are the squeezing amplitude and phase \cite{PhysRevA.99.021801,PhysRevLett.60.764,PhysRevLett.56.1917,PhysRevA.39.2519,PhysRevLett.65.1419,PhysRevX.7.041011,PhysRevApplied.9.044023,Qiu2023}.
	
	\emph{Transient sensing.}---The magnetic-field signal is transduced to the cavity field through magnon-photon coupling and can therefore be detected in the output spectrum. In conventional linear sensing, the Wiener-Khinchin theorem yields a steady-state spectrum in the infinite-measurement-time limit, which is independent of initial conditions \cite{RevModPhys.82.1155}. This description, however, is not sufficient when the signal duration is comparable to the measurement time. In that case, the stationarity assumption breaks down, and the transient dynamics may carry useful information that is lost in the steady-state limit. Although related approaches have been developed for force detection \cite{PhysRevA.64.051401,PhysRevA.65.063803,PhysRevA.102.053515}, nonstationary sensing has not, to our knowledge, been investigated in cavity-magnon platforms. To demonstrate our nonstationary sensing strategy, we consider an impulsive external magnetic field, $\mathbf{B}(t)=\mathbf{B}_0\delta(t)$, where $\mathbf{B}_0$ denotes the time-integrated pulse area \cite{oppenheim1997signals}. From Eq.~(\ref{eq:HL}), one finds that the amplitude quadrature of the cavity field ($\hat a_x$) encodes the $x$- and $z$-components of the magnetic field, whereas the phase quadrature ($\hat a_p$) encodes the $y$- and $z$-components.
	
	To initiate the sensing protocol, the system is first prepared in a suitable initial state and then interacts with the target signal. Here, the initial state is chosen as the steady state of the system in the presence of the injected squeezed environment but in the absence of the magnetic-field signal. This Gaussian state is fully characterized by its covariance matrix (see SM, Sec.~II \cite{SM}). Although a deliberately prepared nonstationary initial state could, in principle, further enhance transient sensing, it would require an immediate switch to the sensing stage after state preparation, which is experimentally unfavorable when the signal arrival time is not precisely known; a detailed discussion is given in SM, Sec.~III \cite{SM}. Based on Eq.~(\ref{eq:HL}), incorporating the input-output relations and the Fourier transform over a finite measurement time $t_m$, we obtain the equivalent input magnetic-field noise power spectral density for sensing the $x(y)$ component as
	\begin{equation}
		S_n^{x(y)}(\omega)
		=
		\frac{
			\left\langle
			\left\{
			\hat u_{x(y)},
			\hat u_{x(y)}^\dagger
			\right\}
			\right\rangle_s
		}{
			4\epsilon_B^2 t_m
		}
		+
		\frac{\left(\frac12+\bar n_a\right)}{2\epsilon_B^2}
		\left[
		\kappa_m+\mathcal K(\omega)\Xi_{x(y)}
		\right],
		\label{eq:Sn}
	\end{equation}
	where $\{,\}$ denotes the anticommutator,
	$\hat u_{x(y)}(\omega)\equiv
	\hat m_{p(x)}
	\mp\gamma B_z^0\hat m_{x(p)}
	\pm\alpha(\omega)\hat a_{x(p)}$,
	$\alpha(\omega)\equiv[-i\omega+(1/2t_m)+(\kappa_m/2)]/g_{am}$,
	$\Xi_{x(y)}\equiv\cosh(2r)\pm\cos(\theta)\sinh(2r)$, and
	\begin{equation}
		\mathcal K(\omega)
		\equiv
		\frac{
			\left|
			4g_{am}^2
			+
			\left(
			\frac1{t_m}-2i\omega-\kappa_a
			\right)
			\left(
			\frac1{t_m}-2i\omega+\kappa_m
			\right)
			\right|^2
		}{
			16g_{am}^2\kappa_a
		}.
		\label{eq:K}
	\end{equation}
	Here, we assume equal thermal occupations for the cavity and magnon reservoirs, $\bar n_a=\bar n_m\equiv\bar n$. Thus, one can evaluate the nonstationary sensing SNR of the $x(y)$-component of the magnetic field, which is defined as
	\begin{equation}
		R_{\mathrm{NSNR}}
		=
		\sqrt{
			S_i^{x(y)}(\omega)
			/S_n^{x(y)}(\omega)
		},
		\label{eq:NSNR}
	\end{equation}
	with
	$S_i^{x(y)}(\omega)=[B_{0x(y)}]^2\Delta f_{\mathrm{ref}}$
	denoting the equivalent signal power spectral density of the externally applied $x(y)$-component magnetic-field pulse, where $\Delta f_{\mathrm{ref}}=1~\mathrm{Hz}$ is the reference measurement bandwidth. This definition converts the squared time-integrated pulse area into a quantity with the same units as $S_n^{x(y)}(\omega)$. The same reference-bandwidth convention is used for both the nonstationary and stationary SNRs. From Eq.~(\ref{eq:Sn}), one can see that the second term gives the stationary contribution, while the first term is the transient contribution determined by the initial-state quadrature covariances. This transient term is the key resource of the present protocol: by engineering the initial Gaussian state, one can reduce the relevant quadrature noise and enhance the finite-time SNR.
	
\begin{figure}
		\includegraphics[width=\columnwidth]{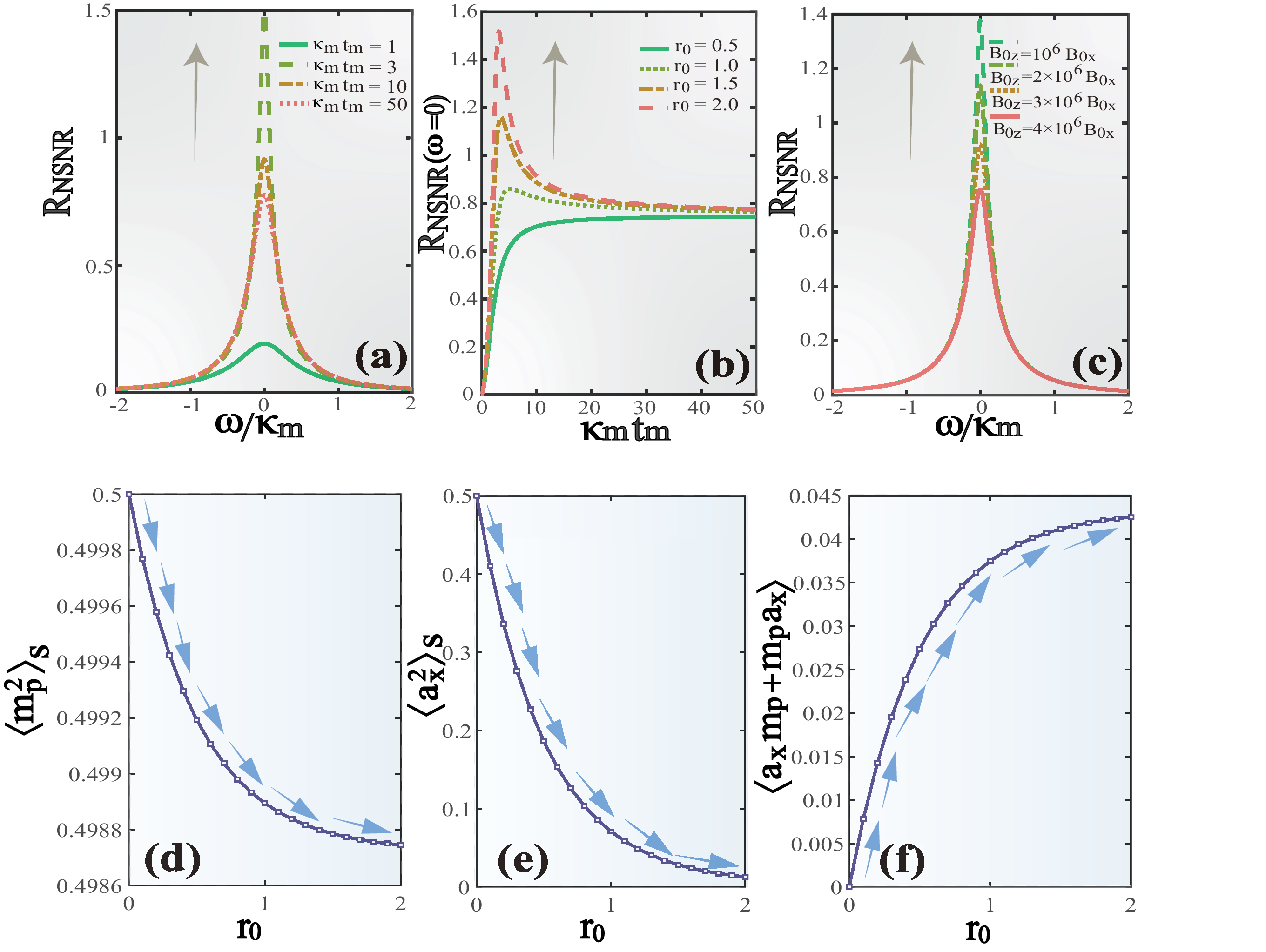}
		\caption{(a) Signal-to-noise ratio $R_{\mathrm{NSNR}}$ versus normalized frequency $\omega/\kappa_m$ for different measurement times, with the preparation-stage squeezing amplitude fixed at $r_0=2$. (b) Resonant SNR at $\omega=0$ versus measurement time for different values of $r_0$. (c) $R_{\mathrm{NSNR}}$ versus $\omega/\kappa_m$ for different longitudinal-field amplitudes, with $r_0=2$ and $\kappa_m t_m=3$. (d) $\langle m_p^2\rangle_s$, (e) $\langle a_x^2\rangle_s$, and (f) $\langle a_xm_p+m_pa_x\rangle_s$ versus $r_0$, with $\theta_0=\pi$. Here, $r_0$ and $\theta_0$ are the squeezing parameters used during state preparation; no squeezing is applied during sensing, i.e., $r=\theta=0$.}
		\label{Fig2}
	\end{figure}

	Here, we present an explicit demonstration of sensing the $x$- and $z$-components of the magnetic field using the amplitude quadrature $\hat a_x$, while the analysis of the phase quadrature is deferred to the SM, Sec.~II \cite{SM}. We adopt experimentally accessible parameters $\omega_{a(m)}/2\pi=7.875~\mathrm{GHz}$, $T=5~\mathrm{mK}$, $\kappa_a/2\pi=2.09~\mathrm{MHz}$, $\kappa_m/2\pi=6~\mathrm{MHz}$, and $g_{am}/(2\pi)=1.77\times10^5~\mathrm{Hz}$ \cite{PhysRevLett.113.083603,PhysRevLett.113.156401,doi:10.1126/sciadv.1501286}. The dependence of the SNR on key parameters is shown in Fig.~\ref{Fig2}. Panel (a) shows that preparation-stage squeezing substantially enhances the SNR at short interrogation times, yielding nearly a twofold improvement at $\kappa_m t_m=3$ compared with the long-time case, $\kappa_m t_m=50$. As shown in panel (b), increasing the squeezing amplitude $r_0$ during the steady-state preparation stage further improves the finite-time performance. Panels (d)--(f) show the relevant covariance elements: the enhancement is dominated by the reduction of $\langle a_x^2\rangle_s$, while the symmetrized cross correlation $\langle \hat a_x\hat m_p+\hat m_p\hat a_x\rangle_s$ remains negligible over the explored parameter range. The nonmonotonic SNR dependence on $\kappa_m t_m$ reflects the competition between signal accumulation and decoherence. At very short times, the field-induced response is still weak; at long times, dissipation washes out the initially prepared squeezing, and the signal accumulation saturates. The optimal interrogation time is therefore of the order of the magnon lifetime, $t_m=O(\kappa_m^{-1})$, where the accumulated signal and the remaining quantum correlations are best balanced. Panel (c) further shows that the SNR for sensing the $x$ component remains above unity even when the $z$ component reaches values as large as $10^6 B_x^0$, indicating strong robustness against longitudinal magnetic fields. This robustness preserves the sensitivity to $B_x$ even in the presence of a large longitudinal component. Simultaneously, $B_z$ produces small but measurable changes in the transient noise spectrum of the filtered amplitude quadrature. Comparing the measured spectrum with a precalibrated reference therefore allows $B_z$ to be extracted quantitatively.
	
	In addition, as discussed in Sec.~II of the SM \cite{SM}, the $y$ component is encoded in the complementary magnon quadrature and can be read out through the cavity phase quadrature. This enables independent detection of the $x$- and $y$-components of the magnetic field. To probe the $y$-component, the preparation-stage squeezing phase $\theta_0$ is set to zero, which optimizes the sensor performance along the $y$-direction. Accordingly, the initial Gaussian state should exhibit reduced noise in the cavity quadrature carrying the target component, namely, $\hat a_x$ for $B_x$ and $\hat a_p$ for $B_y$. The protocol therefore depends only on the relevant second-order quadrature correlations, rather than on a specific pure state. Such squeezed microwave quadratures can be prepared with Josephson parametric amplifiers, while strong cavity-magnon hybridization with YIG spheres has been experimentally demonstrated \cite{Zhang2015,PhysRevLett.113.083603,PhysRevLett.113.156401,Qiu2023}. Building on this capability, we next demonstrate the reconstruction of the full magnetic-field vector and its orientation.
	
	\begin{figure}[htbp]
		\centering
		\includegraphics[width=\columnwidth]{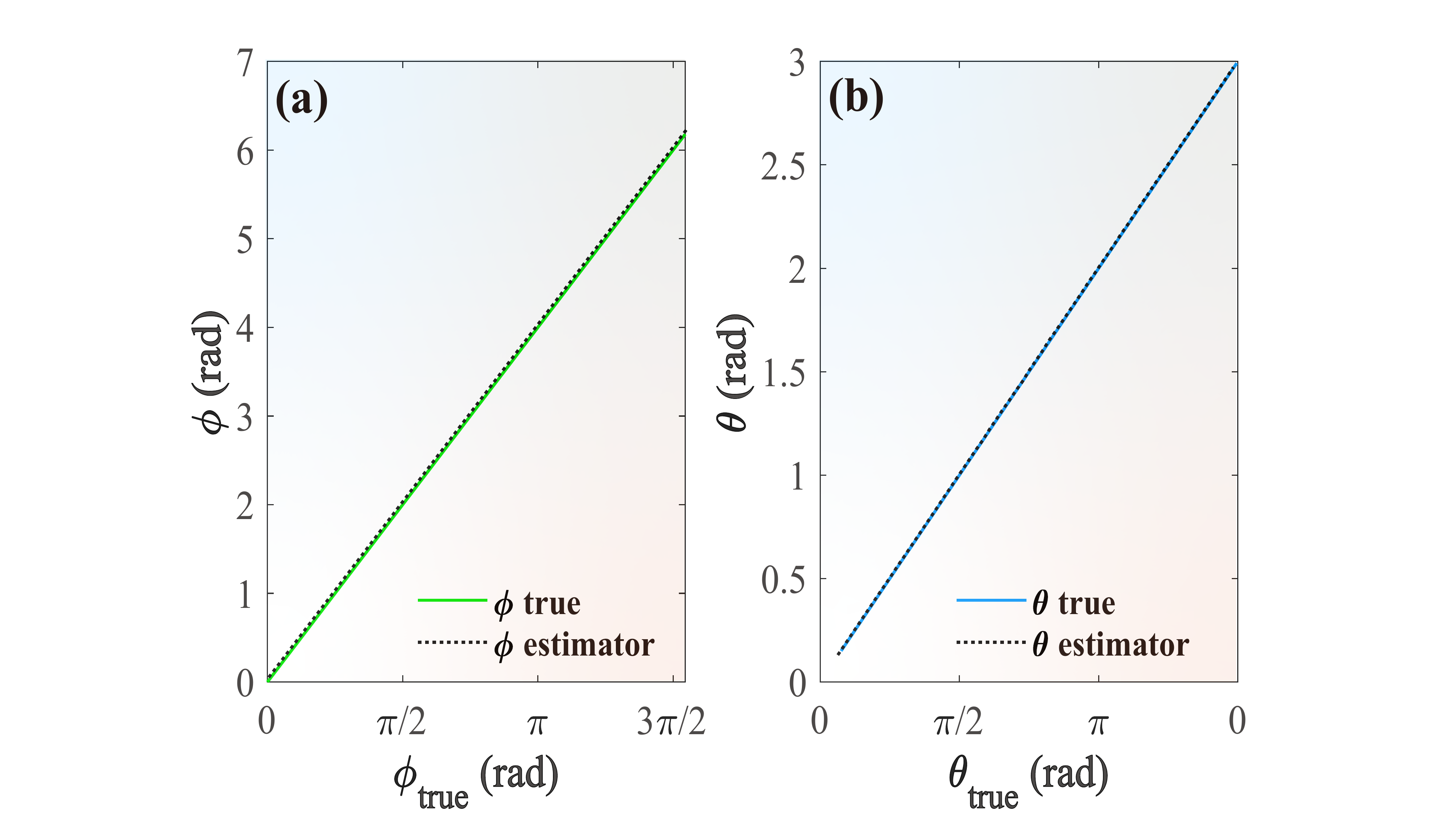}
		\caption{The reconstructed azimuthal angle ($\phi$) (a) and polar angle ($\theta$) (b) of a weak magnetic field using the double-difference estimator. Solid lines denote true angles, and dashed lines denote estimated values.}
		\label{FIG3}
	\end{figure}
	
	Figure~\ref{FIG3} demonstrates the reconstruction of the magnetic-field orientation using a double-difference estimator based on a central finite-difference construction. Both the azimuthal and polar angles closely track their true values, showing that the magnetic-field direction can be determined from the measured output spectra. The transverse components are obtained from frequency-independent signal contributions to the power spectrum, while the longitudinal component is inferred from transient modifications of the noise spectrum. By applying a small calibrated longitudinal bias and double-difference estimator, all stationary noise and transverse contributions are eliminated, yielding a response linear in the longitudinal field amplitude. After subtracting this contribution, the remaining spectra provide the transverse amplitudes, with their signs fixed by calibrated reference pulses. Together, this procedure enables complete reconstruction of the three-dimensional magnetic-field vector, including its magnitude and orientation. Details of the reconstruction protocol are provided in SM, Sec.~IV \cite{SM}.
	
	\emph{Steady-state limit.}---In the long-time limit, the transient part of Eq.~(\ref{eq:Sn}) vanishes, and only the stationary noise spectrum remains. For the impulsive longitudinal signal considered here, $B_z$ no longer affects the long-time output spectrum; hence, a purely steady-state readout cannot reconstruct the full magnetic-field vector. This contrasts directly with the transient protocol, in which $B_z$ is encoded in finite-time changes of the noise spectrum.
	
	By selecting the demodulation phase, one can selectively detect the transverse components $B_x$ or $B_y$. Without injected squeezing, the stationary spectrum can be written as
	\begin{equation}
		S_{B_{x(y)}}^{\mathrm{sym}}(\omega)
		=
		S_{B_{x(y)}}(\omega)
		+
		S_m^s(\omega)
		+
		S_{\mathrm{cavity}}^s(\omega),
		\label{eq:steady}
	\end{equation}
	where $S_{B_{x(y)}}(\omega)$ represents the normalized external magnetic-field signal, $S_m^s(\omega)=\frac{\kappa_m}{2\epsilon_B^2}\left(\bar n_m+\frac12\right)$ is the magnon input noise, and $S_{\mathrm{cavity}}^s(\omega)=\frac{|A|^2}{\kappa_a\epsilon_B^2}\left(\bar n_a+\frac12\right)$ is the cavity-added noise, where
	\[
	A=
	\frac{1}{\sqrt{2}g_{am}}
	\left[
	(\omega-i\kappa_a/2)(\omega+i\kappa_m/2)-g_{am}^2
	\right].
	\]
	As such, one can define the noise ratio as
	\begin{align}
		S_r(\omega)
		&\equiv
		\frac{S_{\mathrm{cavity}}^s(\omega)}{S_m^s(\omega)}
		\nonumber\\
		&=
		\frac{
			(\omega^2+\kappa_a^2/4)(\omega^2+\kappa_m^2/4)
		}{
			g_{am}^2\kappa_a\kappa_m
		}
		+
		\frac{g_{am}^2-2\omega^2}{\kappa_a\kappa_m}
		-\frac12.
		\label{eq:Sr}
	\end{align}
	Thus, $S_r<1$ means that the cavity-added noise is suppressed below the magnon probe-noise floor. At resonance, $\omega=0$, one obtains
	\begin{equation}
		S_r(0)
		=
		\frac{
			(4g_{am}^2-\kappa_a\kappa_m)^2
		}{
			16g_{am}^2\kappa_a\kappa_m
		},
		\label{eq:Sr0}
	\end{equation}
	so the cavity-added noise vanishes under the on-resonance matching condition $g_{am}=\sqrt{\kappa_a\kappa_m}/2$.
	
	\begin{figure}
		\includegraphics[width=\columnwidth]{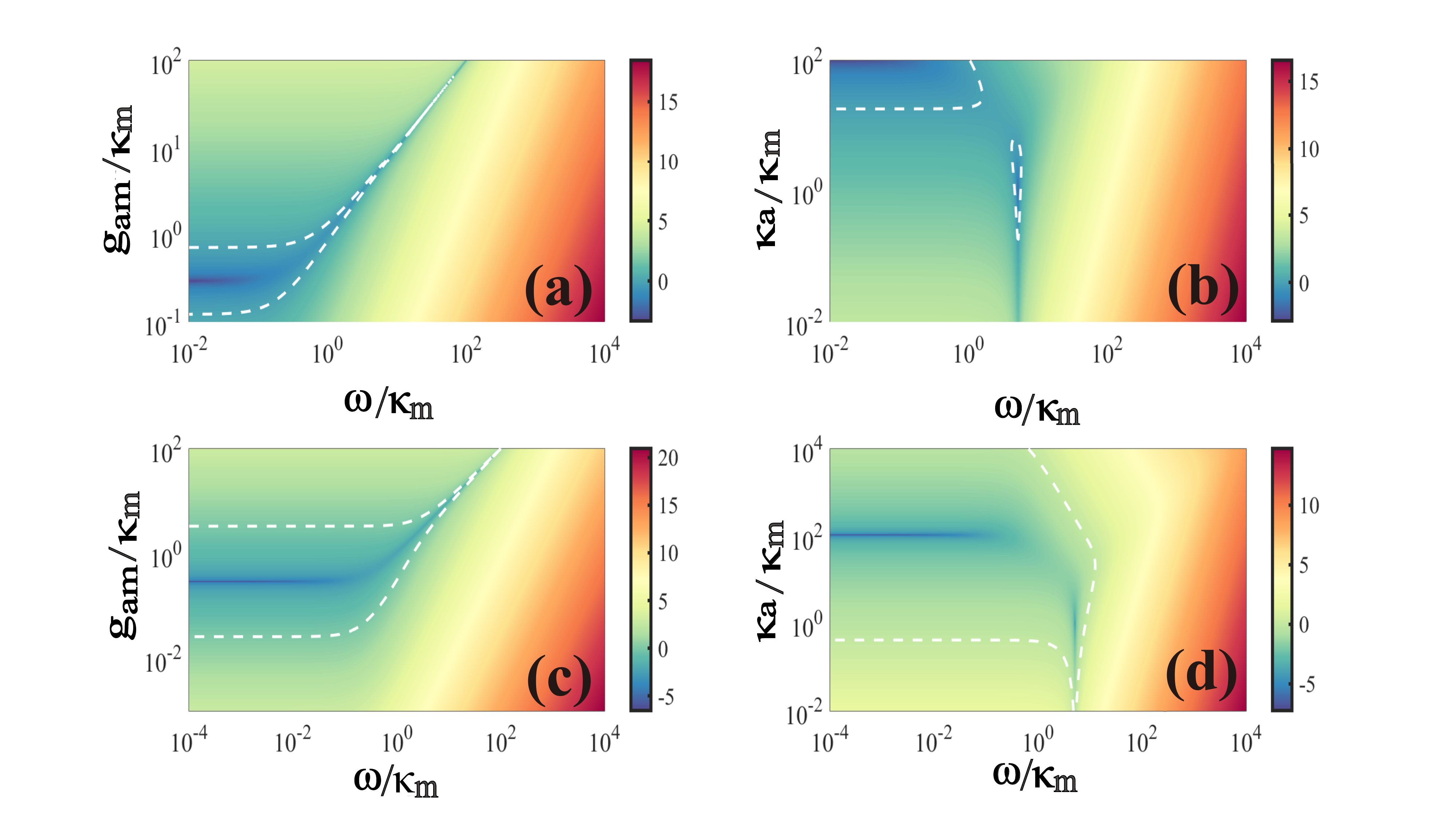}
		\caption{$S_r(\omega)$ versus $\omega/\kappa_m$ and $g_{am}/\kappa_m$ in (a) and (c), and versus $\omega/\kappa_m$ and $\kappa_a/\kappa_m$ in (b) and (d). Panels (a) and (b) show the results without squeezing, whereas panels (c) and (d) include squeezing. The white dashed contour denotes $S_r=1$, separating the region in which the cavity-added noise is below the magnon probe-noise floor from that in which it exceeds this floor.}
		\label{Fig4}
	\end{figure}
	
	Away from resonance, complete cancellation is no longer possible, as shown in Figs.~\ref{Fig4}(a) and \ref{Fig4}(b). If the cavity is driven by a squeezed thermal reservoir, the cavity-added noise becomes
	\begin{equation}
		S_{\mathrm{cavity}}^{\mathrm{sq}}(\omega)
		=
		\left[
		\cosh(2r)\pm\cos(\theta)\sinh(2r)
		\right]
		S_{\mathrm{cavity}}^s(\omega),
		\label{eq:sq}
	\end{equation}
	where the plus and minus signs correspond to sensing $B_x$ and $B_y$, respectively. Figures~\ref{Fig4}(c) and \ref{Fig4}(d) show that squeezing expands the parameter region in which the cavity-added noise lies below the magnon probe-noise floor. The resulting stationary SNR and temperature robustness are discussed in the End Matter.
	
	\emph{Discussion and conclusions.}---We have proposed a transient vector quantum magnetometry protocol for cavity-magnon systems. Starting from an engineered squeezed steady state, finite-time dynamics retain the benefit of squeezing-induced quadrature noise reduction and enhance the short-time SNR beyond conventional unsqueezed steady-state sensing. Quadrature-resolved measurements of the cavity output enable crosstalk-free reconstruction of all three components of a transient magnetic field, thereby extending cavity-magnon magnetometry from scalar detection to vector-resolved signal characterization. This capability is relevant for pulsed spin excitations, nanoscale current transients, magnetic textures, hybrid quantum devices, and weak biomagnetic signals, where the field orientation carries essential information. In the steady-state limit, the cavity-added noise can be canceled under the on-resonance noise-cancellation condition, while injected squeezing suppresses the cavity-added noise away from resonance and broadens the low-noise bandwidth. As shown in the End Matter, an array of $N$ YIG spheres further reduces the magnon-probe noise contribution by a factor of $1/N$ through a collective bright mode. Since cavity-magnon hybridization, coherent microwave driving, and squeezed microwave fields have been experimentally demonstrated \cite{PhysRevLett.113.083603,PhysRevLett.113.156401,PhysRevLett.124.171801,Qiu2023}, our scheme provides a feasible route toward transient, vector-resolved quantum magnetometry.
	
	\emph{Acknowledgments.}---This work was supported by the National Natural Science Foundation of China under Grant No.~12575009.
	
	\emph{Data availability.}---The data supporting the findings of this article are not publicly available but may be obtained from the authors upon reasonable request.
	\bibliography{magnetometry}
	\section{End Matter}
\begin{figure}[htbp]
		\centering
		\includegraphics[width=\columnwidth]{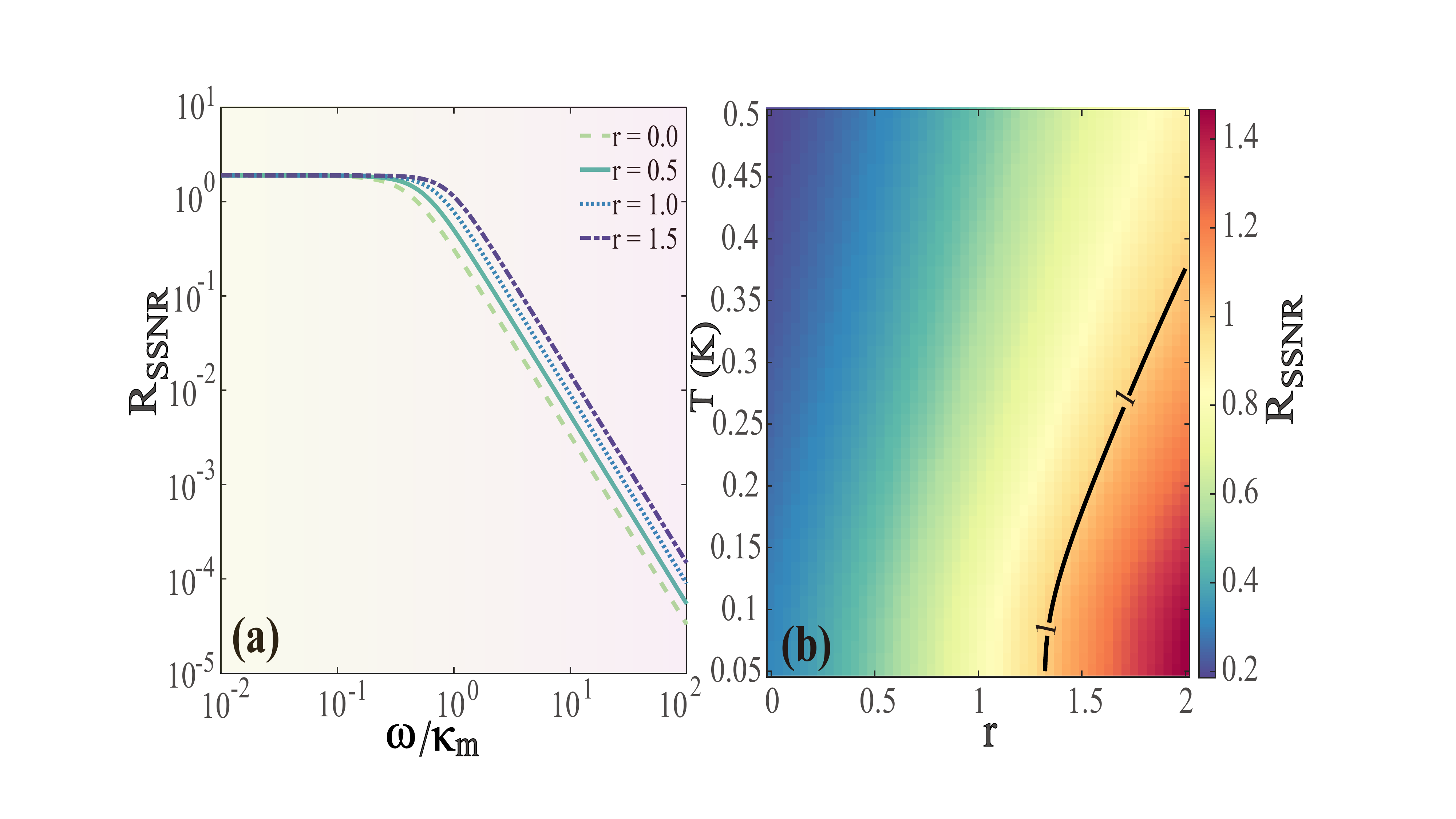}
		\caption{(a) Stationary SNR versus normalized frequency
			$\omega/\kappa_m$ for different squeezing parameters.
			(b) Stationary SNR as a function of temperature and the
			squeezing parameter. The black solid contour denotes the
			threshold $R_{\mathrm{SSNR}}=1$.}
		\label{Fig5}
	\end{figure}

		\paragraph{Stationary SNR and temperature robustness---}
	For comparison with the steady-state noise spectra, we define the stationary SNR as
	\begin{equation}
			R_{\mathrm{SSNR}}
		=\sqrt{\frac{	S_i^{x(y)}(\omega)}{S_{B_{x(y)}}^{\mathrm{sym}}(\omega)\big|_{B_{x(y)}=0}}}.
		\tag{A1}
	\end{equation}
Here, the same reference-bandwidth convention as in Eq.~(\ref{eq:NSNR}) is used, with $\Delta f_{\mathrm{ref}}=1~\mathrm{Hz}$ and $S_i^{x(y)}(\omega)=[B_{0x(y)}]^2\Delta f_{\mathrm{ref}}$. We take $S_i^{x(y)}(\omega)=10^{-24}~\mathrm{T}^2/\mathrm{Hz}$ for the weak magnetic-field pulse, corresponding to an equivalent signal amplitude spectral density of $1~\mathrm{pT}/\sqrt{\mathrm{Hz}}$. Since the signal and noise spectra refer to the same measurement bandwidth, the common bandwidth factor cancels in the SNR ratio. As shown in Fig.~\ref{Fig5}(a), at the optimal coupling strength, the SNR near $\omega\to0$ is almost insensitive to the squeezing parameter, reflecting cancellation of the cavity-added noise. Away from resonance, larger squeezing enhances the SNR over a broader frequency range. Figure~\ref{Fig5}(b) further shows that squeezing permits operation at higher temperatures and improves the achievable SNR.
	\paragraph{Collective enhancement with multiple YIG spheres---}
	We further consider $N$ identical YIG spheres coupled to the same cavity mode. The cavity then interacts with the collective bright magnon mode, leading to scalable suppression of the magnon-noise contribution. Following the same procedure, the stationary noise spectrum without squeezing becomes
	\begin{equation}
		S_{B_{x(y)}}^{\mathrm{sym}\prime}(\omega)
		=
		S_{B_{x(y)}}(\omega)
		+
		\frac{1}{N}S_m^s(\omega)
		+
		\frac{|A'|^2}{N^2|A|^2}
		S_{\mathrm{cavity}}^s(\omega),
		\tag{B1}
	\end{equation}
	where
	\begin{equation}
		A'
		=
		\frac{1}{\sqrt{2}g_{am}}
		\left[
		(\omega-i\kappa_a/2)(\omega+i\kappa_m/2)
		-Ng_{am}^2
		\right].
		\tag{B2}
	\end{equation}
	The magnon-noise contribution is reduced by a factor of $1/N$. At $\omega=0$, the cavity-added noise is fully suppressed when
	\begin{equation}
		\kappa_a\kappa_m
		=
		4Ng_{am}^2.
		\tag{B3}
	\end{equation}
	In this regime, the SNR is limited only by the reduced magnon-noise contribution, which decreases with increasing $N$.
	
	In the off-resonant regime, we denote the optimized noise ratio by $S_N^r(\omega)$. At the optimized coupling, one obtains
	\begin{equation}
		S_N^r(\omega)\big|_{g_{am}^{(\mathrm{opt})}}
		=
		\frac{
			\omega^2\left[
			4\omega^2+(\kappa_a-\kappa_m)^2
			\right]
		}{
			\kappa_a^2\kappa_m^2
		}.
		\tag{B4}
	\end{equation}
	The condition $S_N^r(\omega)\big|_{g_{am}^{(\mathrm{opt})}}<1$ yields
	\begin{equation}
		|\omega|
		<
		\sqrt{
			\frac{
				-(\kappa_a-\kappa_m)^2
				+
				\sqrt{
					(\kappa_a-\kappa_m)^4
					+
					16\kappa_a^2\kappa_m^2
				}
			}{
				8
			}
		}.
		\tag{B5}
	\end{equation}
	For $\kappa_a=\kappa_m=\kappa$, this reduces to
	\begin{equation}
		|\omega|
		<
		\frac{\kappa}{\sqrt{2}}.
		\tag{B6}
	\end{equation}
	Thus, collective coupling reduces the probe-noise floor, while the dissipation rates determine the frequency range over which the cavity-added noise remains subdominant.
\end{document}